\documentclass[fleqn,twoside]{article}
\usepackage{espcrc2}

\usepackage{graphicx}

\newcommand{\AmS}{{\protect\the\textfont2
  A\kern-.1667em\lower.5ex\hbox{M}\kern-.125emS}}

\newcommand{\gsim}{\,\raise0.5ex\hbox{$>$\kern-0.75em\raise-1.1ex\hbox{$\sim$}}\;}
\newcommand{\lsim}{\,\raise0.5ex\hbox{$<$\kern-0.75em\raise-1.1ex\hbox{$\sim$}}\;}

\title{Scalar meson exchange in $\Phi\rightarrow P^0 P^0\gamma$ decays
\thanks{UAB--FT--534 report. 
            To appear in the proceedings of the Photon 2003
             International Conference on the Structure and
             Interactions of the Photon,
             7-11th April 2003 Frascati (Italy).}}

\author{R. Escribano
\address{Grup de F\'{\i}sica Te\`orica and IFAE,
               Universitat Aut\`onoma de Barcelona,\\
               E-08193 Bellaterra (Barcelona), Spain}%
\thanks{Work partly supported by the
             Ministerio de Ciencia y Tecnolog\'{\i}a and FEDER, FPA2002-00748,
             and the EU, HPRN-CT-2002-00311, EURIDICE network.}}
       
\begin{document}
    
\begin{abstract}
The scalar meson exchange in $\phi\rightarrow P^0 P^0\gamma$ decays is 
discussed in a chiral invariant framework where the scalar meson poles 
are incorporated explicitly.
$\phi\rightarrow\pi^0\pi^0\gamma$ and $\phi\rightarrow\pi^0\eta\gamma$ 
are in agreement with recent experimental data and can be 
used to extract valuable information on the properties of 
$f_{0}(980)$ and $a_{0}(980)$ respectively.
$\phi\rightarrow K^0\bar K^0\gamma$ is shown not to pose a background 
problem for testing $CP$ violation at Da$\Phi$ne.
The ratio $\phi\rightarrow f_{0}\gamma/a_{0}\gamma$ is also predicted.
\end{abstract}

\maketitle

\section{INTRODUCTION}
The radiative decays of light vector mesons $(V=\rho,\omega,\phi)$
into a pair of neutral pseudoscalars $(P=\pi^0,K^0,\eta)$,
$V\rightarrow P^0P^0\gamma$, are an excellent laboratory for investigating the
nature and extracting the properties of the light scalar meson resonances
$(S=\sigma,a_{0},f_{0})$.
In addition, their study complements other analyses based on central 
production, $D$ and $J/\psi$ decays, etc.
Particularly interesting are the $\phi$ radiative decays, namely
$\phi\rightarrow\pi^0\pi^0\gamma$, $\phi\rightarrow\pi^0\eta\gamma$ and
$\phi\rightarrow K^0\bar K^0\gamma$, which, as we will see, can provide us with
valuable information on the properties of the $f_{0}(980)$, $a_{0}(980)$ and
$\sigma(500)$.
The ratio $\phi\rightarrow f_{0}\gamma/a_{0}\gamma$ can also be used to 
extract relevant information on the scalar mixing angle.

In Sec.~\ref{expdat}, the most recent experimental data on the
$\phi\rightarrow P^0 P^0\gamma$ decays is presented.
Sec.~\ref{theory} is a short review of the approaches used in the literature to study 
these processes emphasizing the different treatments of the scalar 
contribution.
$\phi\rightarrow\pi^0\pi^0\gamma$, $\phi\rightarrow\pi^0\eta\gamma$ and
$\phi\rightarrow K^0\bar K^0\gamma$ are discussed in
Secs.~\ref{pi0pi0}, \ref{pi0eta} and \ref{K0K0bar}, respectively.
The ratio $\phi\rightarrow f_{0}\gamma/a_{0}\gamma$ is discussed in
Sec.~\ref{ratio}.
Concluding remarks are presented in Sec.~\ref{conclusions}.

\section{EXPERIMENTAL DATA}
\label{expdat}
For $\phi\rightarrow\pi^0\pi^0\gamma$, the first measurements of this decay
have been reported by the SND and CMD-2 Collaborations.
For the branching ratio they obtain
$B(\phi\rightarrow\pi^0\pi^0\gamma)=(1.221\pm 0.098\pm 0.061)\times 10^{-4}$
\cite{Achasov:2000ym}
and $(1.08\pm 0.17\pm 0.09)\times 10^{-4}$ \cite{Akhmetshin:1999di}.
More recently, the KLOE Collaboration has measured
$B(\phi\rightarrow\pi^0\pi^0\gamma)=(1.09\pm 0.03\pm 0.05)\times 10^{-4}$
\cite{Aloisio:2002bt}.
In all the cases, the spectrum is clearly peaked at $m_{\pi\pi}\simeq 970$ MeV,
as expected from an important $f_{0}(980)$ contribution.

For $\phi\rightarrow\pi^0\eta\gamma$, the branching ratios measured by the
SND, CMD-2 and KLOE Collaborations are
$B(\phi\rightarrow\pi^0\eta\gamma)=(8.8\pm 1.4\pm 0.9)\times 10^{-5}$
\cite{Achasov:2000ku},
$(9.0\pm 2.4\pm 1.0)\times 10^{-5}$ \cite{Akhmetshin:1999di}, and
$B(\phi\rightarrow\pi^0\eta\gamma)=(8.51\pm 0.51\pm 0.57)\times 10^{-5}$
$(\eta\rightarrow\gamma\gamma)$ and
$(7.96\pm 0.60\pm 0.40)\times 10^{-5}$ $(\eta\rightarrow\pi^+\pi^-\pi^0)$
\cite{Aloisio:2002bs}.
Again, in all these cases, the observed mass spectrum shows a
significant enhancement at large $\pi^0\eta$ invariant mass that is
interpreted as a manifestation of the dominant contribution of the
$a_{0}(980)\gamma$ intermediate state.

For $\phi\rightarrow K^0\bar K^0\gamma$, no experimental data is yet 
available.

For the ratio $\phi\rightarrow f_{0}\gamma/a_{0}\gamma$, the 
experimental value measured by the KLOE Collaboration is
$R(\phi\rightarrow f_{0}\gamma/a_{0}\gamma)=6.1\pm 0.6$ \cite{Aloisio:2002bs}.

\section{THEORETICAL FRAMEWORK}
\label{theory}
A first attempt to explain the $V\rightarrow P^0P^0\gamma$ decays was done in
Ref.~\cite{Bramon:1992kr} using the vector meson dominance (VMD) model.
In this framework, the $V\rightarrow P^0P^0\gamma$ decays proceed through the
decay chain $V\rightarrow VP^0\rightarrow P^0P^0\gamma$.
The intermediate vectors exchanged are $V=V^\prime=\rho$ for 
$\phi\rightarrow\pi^0\pi^0\gamma$, $V=\rho$ and $V^\prime=\omega$ for
$\phi\rightarrow\pi^0\eta\gamma$, and $V=\bar K^{\ast 0}$ and 
$V^\prime=K^{\ast 0}$ for $\phi\rightarrow K^0\bar K^0\gamma$.
The calculated branching ratios
$B_{\phi\rightarrow\pi^0\pi^0\gamma}^{\rm VMD}=1.2\times 10^{-5}$,
$B_{\phi\rightarrow\pi^0\eta\gamma}^{\rm VMD}=5.4\times 10^{-6}$ and
$B_{\phi\rightarrow K^0\bar K^0\gamma}^{\rm VMD}=2.7\times 10^{-12}$
\cite{Bramon:1992kr}
are found (for the first two processes) to be substantially smaller than the
experimental results.

Later on, the $V\rightarrow P^0P^0\gamma$ decays were studied in a 
Chiral Perturbation Theory (ChPT) context enlarged to included on-shell vector
mesons \cite{Bramon:1992ki}. In this formalism,
$B_{\phi\rightarrow\pi^0\pi^0\gamma}^\chi=5.1\times 10^{-5}$,
$B_{\phi\rightarrow\pi^0\eta\gamma}^\chi=3.0\times 10^{-5}$ and
$B_{\phi\rightarrow K^0\bar K^0\gamma}^\chi=7.6\times 10^{-9}$.
Taking into account both chiral and VMD contributions, one finally obtains
$B_{\phi\rightarrow\pi^0\pi^0\gamma}^{{\rm VMD}+\chi}=6.1\times 10^{-5}$ 
and
$B_{\phi\rightarrow\pi^0\eta\gamma}^{{\rm VMD}+\chi}=3.6\times 10^{-5}$,
which are still below the experimental results, and
$B_{\phi\rightarrow K^0\bar K^0\gamma}^{{\rm VMD}+\chi}=7.6\times 10^{-9}$
\cite{Bramon:1992ki}.

Additional contributions are thus certainly required and the most natural 
candidates are the contributions coming from the exchange of scalar resonances.
A first model including the scalar resonances explicitly is the
\emph{no structure model}, where the $V\rightarrow P^0P^0\gamma$ decays
proceed through the decay chain
$V\rightarrow S\gamma\rightarrow P^0P^0\gamma$
and the coupling $VS\gamma$ is considered as pointlike.
This model is ruled out by experimental data on
$\phi\rightarrow\pi^0\pi^0\gamma$ decays \cite{Achasov:2000ym}.
A second model is the \emph{kaon loop model} \cite{Achasov:1987ts},
where the initial vector decays into a pair of charged kaons that,
after the emission of a photon, rescatter into a pair of neutral pseudoscalars
through the exchange of scalar resonances.

The previous two models include the scalar resonances {\it ad hoc},
and the pseudoscalar rescattering amplitudes are not chiral invariant.
This problem is solved in the next two models which are based not only on the
\emph{kaon loop model} but also on chiral symmetry.
The first one is a chiral unitary approach (U$\chi$) where the 
scalar resonances are generated dynamically by unitarizing the one-loop
pseudoscalar amplitudes.
In this approach,
$B_{\phi\rightarrow\pi^0\pi^0\gamma}^{\rm U\chi }=8\times 10^{-5}$,
$B_{\phi\rightarrow\pi^0\eta\gamma}^{\rm U\chi}=8.7\times 10^{-5}$
\cite{Marco:1999df} and
$B_{\phi\rightarrow K^0\bar K^0\gamma}^{\rm U\chi}=5\times 10^{-8}$
\cite{Oller:1998ia}.
The second model is the Linear Sigma Model (L$\sigma$M), a well-defined
$U(3)\times U(3)$ chiral model which incorporates {\it ab initio} the
pseudoscalar and scalar mesons nonets.
The advantage of the L$\sigma$M is to incorporate explicitly the effects of
scalar meson poles while keeping the correct behaviour at low invariant masses
expected from ChPT.

In the next four sections, we discuss the scalar contributions to the
$\phi\rightarrow\pi^0\pi^0\gamma$, $\phi\rightarrow\pi^0\eta\gamma$ and
$\phi\rightarrow K^0\bar K^0\gamma$ decays,
and the ratio $\phi\rightarrow f_{0}\gamma/a_{0}\gamma$,
in the framework of the L$\sigma$M.

\section{$\phi\rightarrow\pi^0\pi^0\gamma$}
\label{pi0pi0}
The scalar contribution to this process is driven by the decay chain
$\phi\rightarrow K^+K^-(\gamma)\rightarrow\pi^0\pi^0\gamma$.
The contribution from pion loops is known to be negligible due to the Zweig rule.
The amplitude for 
$\phi(q^\ast,\epsilon^\ast)\rightarrow\pi^0(p)\pi^0(p^\prime)\gamma(q,\epsilon)$
is given by \cite{Bramon:2002iw}
\begin{equation}
\label{Aphipi0pi0gamma}
{\cal A}=\frac{eg_{s}}{2\pi^2 m^2_{K^+}}\,\{a\}\,L(m^2_{\pi^0\pi^0})
\times{\cal A}_{K^+ K^-\rightarrow\pi^0\pi^0}^{\mbox{\scriptsize 
L$\sigma$M}}\, ,
\end{equation} 
where
$\{a\}=(\epsilon^\ast\cdot\epsilon)\,(q^\ast\cdot q)-
           (\epsilon^\ast\cdot q)\,(\epsilon\cdot q^\ast)$, 
$m^2_{\pi^0\pi^0}\equiv s$ is the dipion invariant mass and $L(m^2_{\pi^0\pi^0})$
is a loop integral function.
The $\phi K\bar K$ coupling constant $g_{s}$ takes the value $|g_s|\simeq 4.5$
to agree with $\Gamma_{\phi\rightarrow K^+K^-}^{\rm exp}= 2.10$ MeV
\cite{Hagiwara:fs}.
The $K^+ K^-\rightarrow\pi^0\pi^0$ amplitude in Eq.~(\ref{Aphipi0pi0gamma})
is calculated using the L$\sigma$M and turns out to be
\begin{equation}
\label{AKKpipiLsM}
\begin{array}{l}
{\cal A}_{K^+K^-\rightarrow\pi^0\pi^0}^{\mbox{\scriptsize L$\sigma$M}}=
\frac{m^2_{\pi}-s/2}{2f_\pi f_K}\\[1ex]
\quad+\frac{s-m^2_{\pi}}{2f_\pi f_K}
\times\left[
\frac{m^2_K-m^2_{\sigma}}{D_{\sigma}(s)}
{\rm c}\phi_S({\rm c}\phi_S-\sqrt{2}\,{\rm s}\phi_S)\right.\\
\qquad\qquad\quad\left.+
\frac{m^2_K-m^2_{f_0}}{D_{f_0}(s)}
{\rm s}\phi_S({\rm s}\phi_S+\sqrt{2}\,{\rm c}\phi_S)
\right]\, ,
\end{array}
\end{equation}
where $D_{S}(s)$ are the $S=\sigma, f_0$ propagators,
$\phi_S$ is the scalar mixing angle in the quark-flavour basis
and $({\rm c}\phi_S, {\rm s}\phi_S)\equiv (\cos\phi_S, \sin\phi_S)$.
A Breit-Wigner propagator is used for the $\sigma$, while for the $f_{0}$
a complete one-loop propagator taking into account finite width 
corrections is preferable \cite{Achasov:1987ts,Escribano:2002iv}.

It is worth mentioning that the amplitude (\ref{Aphipi0pi0gamma}) is the 
result of adding a resonant amplitude containing the scalar poles in the $s$-channel
---the $\sigma$ and $f_{0}$ poles in Eq.~(\ref{AKKpipiLsM})---
and a non-resonant one including all other contributions.
This latter is obtained from the subtraction
\begin{equation}
\label{Anonresonant}
\begin{array}{l}
{\cal A}_{\phi\rightarrow\pi^0\pi^0\gamma}^{\mbox{\scriptsize non-res.}}=
{\cal A}_{\phi\rightarrow\pi^0\pi^0\gamma}^{\chi}\\[1ex]
\qquad
-\lim_{m_{\sigma, f_{0}}\rightarrow\infty}
{\cal A}_{\phi\rightarrow\pi^0\pi^0\gamma}^{\mbox{\scriptsize res.}}
\propto\frac{m^2_{\pi}-s/2}{2f_\pi f_K}\ ,
\end{array}
\end{equation}
where ${\cal A}_{\phi\rightarrow\pi^0\pi^0\gamma}^{\chi}$ 
is the corresponding chiral-loop amplitude,
${\cal A}_{\phi\rightarrow\pi^0\pi^0\gamma}^{\chi}\propto s/4f_{\pi}f_{K}$,
and ${\cal A}_{\phi\rightarrow\pi^0\pi^0\gamma}^{\mbox{\scriptsize res.}}$
is the aforementioned resonant contribution.
In this way,
${\cal A}_{\phi\rightarrow\pi^0\pi^0\gamma}^{\mbox{\scriptsize non-res.}}$
encodes the effects of all resonances in the $t$- and $u$-channel
---and also of higher spin resonances in the $s$-channel---
in the limit $m_{R}\rightarrow\infty$.

Notice also that for $m_{S}\rightarrow\infty\ (S=\sigma,f_{0})$,
the amplitude (\ref{Aphipi0pi0gamma}) reduces to the chiral-loop prediction
and is thus expected to account for the lowest part of the $\pi\pi$ spectrum.
In addition, the presence of the scalar propagators in Eq.~(\ref{AKKpipiLsM})
should be able to reproduce the effects of the $f_{0}$ (and the $\sigma$)
pole(s) at higher $\pi\pi$ invariant mass values.
This complementarity between ChPT and the L$\sigma$M makes the whole analysis
quite reliable.

\begin{figure}[t]
\centerline{\includegraphics[width=0.45\textwidth]{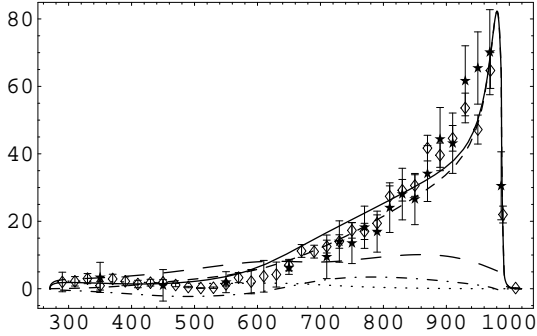}} 
\caption{\small
$dB(\phi\rightarrow\pi^0\pi^0\gamma)/dm_{\pi^0\pi^0} \times 10^8$
(in MeV$^{-1}$) {\it versus} $m_{\pi^0\pi^0}$ (in MeV). 
The dashed, dotted and dot-dashed lines correspond to the contributions from
the L$\sigma$M, VMD and their interference, respectively. 
The solid line is the total result.
The long-dashed line is the chiral-loop prediction.
Experimental data are taken from Ref.~\protect\cite{Achasov:2000ym} (solid star)
and Ref.~\protect\cite{Aloisio:2002bt} (open diamond).}
\label{dBdmpipiallcontrib}
\end{figure}
The final results for ${\cal A}(\phi\rightarrow\pi^0\pi^0\gamma)$ are then the
sum of the L$\sigma$M contribution in Eq.~(\ref{Aphipi0pi0gamma}) plus the VMD
contribution that can be found in Ref.~\cite{Bramon:2002iw}.
The $\pi^0\pi^0$ invariant mass distribution,
with the separate contributions from the L$\sigma$M, VMD and their interference,
as well as the total result, are shown in 
Fig.~\ref{dBdmpipiallcontrib}.
We use $m_\sigma=478$ MeV \cite{Aitala:2001xu}, 
$\Gamma_\sigma=256$ MeV, as required by the L$\sigma$M, 
$m_{f_{0}}=985$ MeV and $\phi_{S}=-9^\circ$
\cite{Bramon:2002iw}.
Notice that the contribution of the $\sigma$ to this process is suppressed
since $g_{\sigma KK}\propto (m^2_{\sigma}-m^2_{K})\simeq 0$ for
$m_{\sigma}\simeq m_{K}$
---by contrast, the chiral loop prediction shows no suppression in the region
$m_{\pi\pi}\simeq 500$ MeV, see Fig.~\ref{dBdmpipiallcontrib}.

Integrating the $\pi^0\pi^0$ invariant mass distribution over the whole physical
region one finally obtains
$B(\phi\rightarrow\pi^0\pi^0\gamma)=1.16\times 10^{-4}$.
The shape of the $\pi\pi$ mass spectrum and the branching ratio are in
agreement with the experimental results.
However, both predictions are very sensitive to the values of the $f_{0}$ mass
and the scalar mixing angle
(this latter because of $g_{f_{0}\pi\pi}\propto\sin\phi_{S}$).
Consequently, the $\phi\rightarrow\pi^0\pi^0\gamma$ decay could be used to 
extract valuable information on these parameters.

\section{$\phi\rightarrow\pi^0\eta\gamma$}
\label{pi0eta}
\begin{figure}[t]
\centerline{\includegraphics[width=0.45\textwidth]{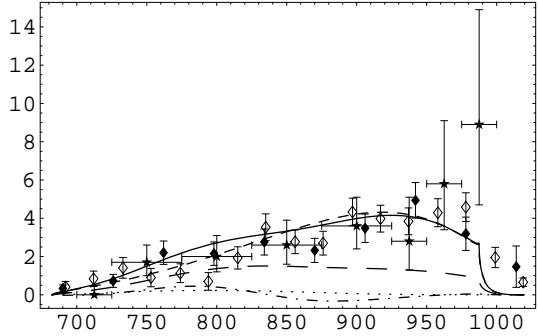}} 
\caption{\small
$dB(\phi\rightarrow\pi^0\eta\gamma)/dm_{\pi^0\eta} \times 10^7$
(in MeV$^{-1}$) {\it versus} $m_{\pi^0\eta}$ (in MeV). 
The curves follow the same conventions as in Fig.~\protect\ref{dBdmpipiallcontrib}.
Experimental data are taken from Ref.~\protect\cite{Achasov:2000ku} (solid star)
and Ref.~\protect\cite{Aloisio:2002bs}:
(open diamond)  from $\eta\rightarrow\gamma\gamma$ and
(solid diamond) from $\eta\rightarrow\pi^+\pi^-\pi^0$.}
\label{dBdmpietaallcontrib}
\end{figure}
The scalar contribution to the $\phi\rightarrow\pi^0\eta\gamma$ decay
is identical to that of $\phi\rightarrow\pi^0\pi^0\gamma$ with the replacement
of ${\cal A}_{K^+K^-\rightarrow\pi^0\pi^0}^{\mbox{\scriptsize L$\sigma$M}}$
by ${\cal A}_{K^+K^-\rightarrow\pi^0\eta}^{\mbox{\scriptsize L$\sigma$M}}$
in Eq.~(\ref{AKKpipiLsM}).
This latter amplitude is written as
\begin{equation}
\label{AKKpietaLsM}
\begin{array}{l}
{\cal A}_{K^+K^-\rightarrow\pi^0\eta}^{\mbox{\scriptsize L$\sigma$M}}
=\frac{m^2_{\eta}+m^2_{\pi}-s}{4f_\pi f_K}
({\rm c}\phi_{P}-\sqrt{2}\,{\rm s}\phi_{P})\\[1ex]
\qquad\qquad\qquad
+\frac{s-m^2_{\eta}}{2f_\pi f_K}
\frac{m^2_K-m^2_{a_0}}{D_{a_0}(s)}\,{\rm c}\phi_{P}\, ,
\end{array}
\end{equation}
where $\phi_{P}$ is the pseudoscalar mixing angle and
$D_{a_0}(s)$ the complete one-loop $a_{0}$ propagator \cite{Achasov:1987ts}.

The separate contributions to the $\pi^0\eta$ invariant mass distribution,
and the total result, are shown in Fig.~\ref{dBdmpietaallcontrib}.
The chiral loop prediction is also included for comparison.
We use $m_{a_{0}}=984.7$ MeV \cite{Hagiwara:fs} and
$\phi_{P}=41.8^\circ$ \cite{Aloisio:2002vm}.
Integrating the $\pi^0\eta$ invariant mass spectrum one obtains
$B(\phi\rightarrow\pi^0\eta\gamma)=8.3\times 10^{-5}$.
The $\pi^0\eta$ mass spectrum and the branching ratio are in fair agreement
with experimental results and with previous phenomenological estimates
\cite{Bramon:2000vu,Escribano:2000fs}.

\section{$\phi\rightarrow K^0\bar K^0\gamma$}
\label{K0K0bar}
This process is interesting to study since it could pose a background 
problem for testing $CP$ violation at Da$\Phi$ne.
The analysis of $CP$-violating decays in $\phi\rightarrow K^0\bar K^0$ has been proposed
as a way to measure the ratio $\epsilon^\prime/\epsilon$ \cite{Dunietz:1986jf}, 
but because this means looking for a very small effect,
a $B(\phi\rightarrow K^0\bar K^0\gamma)\gsim 10^{-6}$
will limit the precision of such a measurement.

The scalar contribution to the $\phi\rightarrow K^0\bar K^0\gamma$ decay
is again identical to that of $\phi\rightarrow\pi^0\pi^0\gamma$
(since it is driven by the same charged kaon loop) with the replacement
of ${\cal A}_{K^+K^-\rightarrow\pi^0\pi^0}^{\mbox{\scriptsize L$\sigma$M}}$
by ${\cal A}_{K^+K^-\rightarrow K^0\bar K^0}^{\mbox{\scriptsize L$\sigma$M}}$
in Eq.~(\ref{AKKpipiLsM}).
This latter amplitude is written as
\begin{equation}
\label{AKKKKKLsM}
\begin{array}{l}
{\cal A}_{K^+K^-\rightarrow K^0\bar K^0}^{\mbox{\scriptsize L$\sigma$M}}=
\frac{m^2_K-s/2}{2f_K^2}-\frac{s-m^2_K}{4f_K^2}\\[1ex]
\quad
\times
\left[
\frac{m^2_K-m^2_{a_0}}{D_{a_0}(s)}
-\frac{m^2_K-m^2_{f_0}}{D_{f_0}(s)}({\rm s}\phi_S +\sqrt{2}{\rm c}\phi_S)^2
\right.\\[1ex]
\qquad
\left.
-\frac{m^2_K-m^2_{\sigma}}{D_{\sigma}(s)}({\rm c}\phi_S -\sqrt{2}{\rm s}\phi_S)^2
\right]\, ,
\end{array}
\end{equation}
where the contributions of the $a_{0}$, $f_{0}$ and $\sigma$ mesons 
are explicitly shown.
The $\sigma$ contribution to this process is negligible not only because of the
$\sigma KK$ coupling suppression if $m_{\sigma}\simeq m_{K}$
but also for kinematical reasons.
The contributions of the $a_{0}$ and $f_{0}$ are of the same order 
but with opposite sign 
($g_{f_{0}K^+K^-}=g_{f_{0}K^0\bar K^0}$ and $g_{a_{0}K^+K^-}=-g_{a_{0}K^0\bar K^0}$
due to isospin invariance).

\begin{figure}[t]
\centerline{\includegraphics[width=0.45\textwidth]{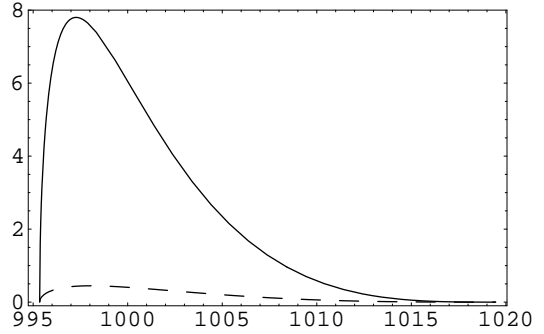}} 
\caption{\small
$dB(\phi\rightarrow K^0\bar K^0\gamma)/dm_{K^0\bar K^0} \times 10^9$
(in MeV$^{-1}$) {\it versus} $m_{K^0\bar K^0}$ (in MeV). 
The solid line is the total result.
The long-dashed line is the chiral-loop prediction.}
\label{dBdmKKbarallcontrib}
\end{figure}

The $K^0\bar K^0$ invariant mass distribution is shown in 
Fig.~\ref{dBdmKKbarallcontrib}.
The chiral loop prediction is also included.
Integrating the $K^0\bar K^0$ invariant mass spectrum one obtains
$B(\phi\rightarrow K^0\bar K^0\gamma)=6\times 10^{-8}$.
This value is in agreement with previous phenomenological estimates
\cite{Oller:1998ia,Close:ay}
(see also Ref.~\cite{Close:ay} for a review of earlier predictions).
Notice that the branching ratio obtained here is one order of magnitude 
larger than the chiral-loop prediction \cite{Bramon:1992ki}.
However, it is still one order of magnitude smaller than the limit, 
${\cal O}(10^{-6})$, in order to pose a background problem for testing 
$CP$-violating decays at Da$\Phi$ne.

\section{$\phi\rightarrow f_{0}\gamma/a_{0}\gamma$}
\label{ratio}
In the kaon loop model, these two processes are driven by the decay 
chain
$\phi\rightarrow K^+K^-(\gamma)\rightarrow f_{0}\gamma$ and
$a_{0}\gamma$.
The amplitudes are given by
\begin{equation}
\label{Aratio}
{\cal A}=\frac{eg_{s}}{2\pi^2 m^2_{K^+}}\,\{a\}\,L(m^2_{f_{0}(a_{0})})
\times g_{f_{0}(a_{0})K^+K^-}\, ,
\end{equation} 
where the scalar coupling constants are fixed within the L$\sigma$M to
\begin{equation}
\label{couplingsratio}
\begin{array}{l}
g_{f_{0}K^+K^-}=\frac{m^2_{K}-m^2_{f_{0}}}{2f_{K}}
({\rm s}\phi_S +\sqrt{2}{\rm c}\phi_S)\, ,\\[1ex]
g_{a_{0}K^+K^-}=\frac{m^2_{K}-m^2_{a_{0}}}{2f_{K}}\, .
\end{array}
\end{equation}
The ratio of the two branching ratios is thus
\begin{equation}
\label{R}
\begin{array}{l}
R_{\phi\rightarrow f_{0}\gamma/a_{0}\gamma}^{\mbox{\scriptsize L$\sigma$M}}=
\frac{|L(m^2_{f_{0}})|^2}{|L(m^2_{a_{0}})|^2}
\frac{\left(1-m^2_{f_{0}}/m^2_{\phi}\right)^3}
       {\left(1-m^2_{a_{0}}/m^2_{\phi}\right)^3}\\[1ex]
\qquad\qquad\quad\ 
       \times\frac{g^2_{f_{0}K^+K^-}}{g^2_{a_{0}K^+K^-}}
\simeq ({\rm s}\phi_S +\sqrt{2}{\rm c}\phi_S)^2\, ,
\end{array}
\end{equation}
where the approximation is valid for $m_{f_{0}}\simeq m_{a_{0}}$.
For $\phi_{S}=-9^\circ$, one gets
$R_{\phi\rightarrow f_{0}\gamma/a_{0}\gamma}^{\mbox{\scriptsize L$\sigma$M}}
 \simeq 1.5$
which should be compared to the experimental value
$R_{\phi\rightarrow f_{0}\gamma/a_{0}\gamma}^{\mbox{\scriptsize KLOE}}
 =6.1\pm 0.6$ \cite{Aloisio:2002bs}.
However, this value is obtained from a large destructive interference 
between the $f_{0}\gamma$ and $\sigma\gamma$ contributions to
$\phi\rightarrow\pi^0\pi^0\gamma$, in disagreement with other 
experiments \cite{Achasov:2000ym}.
Conversely, the measurement of the ratio could be used to get some insight into
the value of the scalar mixing angle.

\section{CONCLUSIONS}
\label{conclusions}
\begin{itemize}
\item
The radiative decays $\phi\rightarrow\pi^0\pi^0\gamma$,
$\phi\rightarrow\pi^0\eta\gamma$ and $\phi\rightarrow K^0\bar K^0\gamma$
have been shown to be very useful to extract relevant information on the
properties of the $f_{0}(980)$, $a_{0}(980)$ and $\sigma(500)$
scalar resonances.

\item
The complementary between ChPT and the L$\sigma$M is used to parametrize
the needed scalar amplitudes.
This guarantees the appropriate behaviour at low dimeson invariant masses
but also allows to include the effects of the scalar meson poles.

\item
The L$\sigma$M predictions for the invariant mass spectra and their
respective branching ratios of the
$\phi\rightarrow\pi^0\pi^0\gamma$ and $\phi\rightarrow\pi^0\eta\gamma$
decays are compatible with experimental data.

\item
The prediction for $\phi\rightarrow\pi^0\pi^0\gamma$ is dominated by
$f_{0}(980)$ exchange and is strongly dependent on the values of $m_{f_{0}}$
and $\phi_{S}$.
For the preferred values $m_{f_{0}}=985$ MeV and $\phi_{S}=-9^\circ$,
one obtains $B(\phi\rightarrow\pi^0\pi^0\gamma)=1.16\times 10^{-4}$.
The suppression of the $\sigma$ contribution to this process is 
explained in terms of the smallness of the $\sigma KK$ coupling for
$m_{\sigma}\simeq m_{K}$.
The process $\phi\rightarrow\pi^0\eta\gamma$ is dominated by $a_{0}(980)$ exchange.
For the values $m_{a_{0}}=984.8$ MeV and $\phi_{P}=41.8^\circ$,
one obtains $B(\phi\rightarrow\pi^0\eta\gamma)=8.3\times 10^{-5}$.

\item
The decay $\phi\rightarrow K^0\bar K^0\gamma$ is confirmed not to pose
a background problem for testing $CP$ violation at Da$\Phi$ne.
The ratio $\phi\rightarrow f_{0}\gamma/a_{0}\gamma$ may be used to 
obtain valuable information on the scalar mixing angle and on the 
nature of the $f_{0}(980)$ and $a_{0}(980)$ scalar states.
\end{itemize}

\section*{Acknowledgements}
I would like to express my gratitude to the Photon 2003
Organizing Committee
(and in particular to G.~Pancheri and A.~Mantella) 
for the opportunity of presenting this contribution,
and for the pleasant and interesting conference we have enjoyed.

\end{document}